\documentclass[aps,reprint,prb,superscriptaddress]{revtex4-1}

\usepackage{ulem}
\usepackage{graphicx}
\usepackage{xcolor}
\usepackage{amsmath}
\usepackage{hyperref} 
\usepackage{makecell}

\def\abinitio{{\it ab~initio}}

  \def\abinitio{\textit{ab\,initio} }
\begin{document}
	\title{Electronic structure study of YNbTiO$_6$ vs.\ CaNb$_2$O$_6$ with U, Pu and minor actinide substitutions using compound-tunable embedding potential method}
	
\author{D.A.\ Maltsev}\email[]{malcev\_da@pnpi.nrcki.ru}
%	\affiliation{Petersburg Nuclear Physics Institute named by B.P.\ Konstantinov of National Research Center ``Kurchatov Institute'' (NRC ``Kurchatov Institute'' - PNPI), 188300, Russian Federation, Leningrad district, Gatchina, mkr.\ Orlova roscha, 1.}
	
\author{Yu.V.\ Lomachuk}
%	\affiliation{Petersburg Nuclear Physics Institute named by B.P.\ Konstantinov of National Research Center ``Kurchatov Institute'' (NRC ``Kurchatov Institute'' - PNPI), 188300, Russian Federation, Leningrad district, Gatchina, mkr.\ Orlova roscha, 1.}
	
\author{V.M.\ Shakhova}
%	\affiliation{Petersburg Nuclear Physics Institute named by B.P.\ Konstantinov of National Research Center ``Kurchatov Institute'' (NRC ``Kurchatov Institute'' - PNPI), 188300, Russian Federation, Leningrad district, Gatchina, mkr.\ Orlova roscha, 1.}

\author{N.S.\ Mosyagin}
%	\affiliation{Petersburg Nuclear Physics Institute named by B.P.\ Konstantinov of National Research Center ``Kurchatov Institute'' (NRC ``Kurchatov Institute'' - PNPI), 188300, Russian Federation, Leningrad district, Gatchina, mkr.\ Orlova roscha, 1.}
	
\author{D.O.\ Kozina}
%	\affiliation{Petersburg Nuclear Physics Institute named by B.P.\ Konstantinov of National Research Center ``Kurchatov Institute'' (NRC ``Kurchatov Institute'' - PNPI), 188300, Russian Federation, Leningrad district, Gatchina, mkr.\ Orlova roscha, 1.}
		
\author{A.V.\ Titov}\email[]{titov\_av@pnpi.nrcki.ru}

 \affiliation{Petersburg Nuclear Physics Institute named by B.P.\ Konstantinov of National Research Center ``Kurchatov Institute'' (NRC ``Kurchatov Institute'' - PNPI), 188300, Russian Federation, Leningrad district, Gatchina, mkr.\ Orlova roscha, 1.}
%   \affiliation{Saint Petersburg State University, 7/9 Universitetskaya nab., 199034 St. Petersburg,  Russia}	
%----------------------------------------------
\date{\today~-nXX-version}
%----------------------------------------------
	
\begin{abstract}

The compound-tunable embedding potential (CTEP) method is applied to study actinide substitutions in the niobate crystals YNbTiO$_6$ and CaNb$_2$O$_6$. Two one-center clusters centered on Ca and Y are built and 20 substitutions of Ca and Y with U, Np, Pu, Am, and Cm in four different oxidation states were made for each cluster. Geometry relaxation is performed for each resulting structure, and electronic properties are analyzed by evaluating the spin density distribution and X-ray emission spectra chemical shifts.
Though the studied embedded clusters with actinides having the same oxidation state are found in general to yield similar local structure distortions, for Am and Cm in high ``starting'' oxidation states the electron transfer from the environment was found, resulting in decrease of their oxidation states, while for ``starting'' U$^{\rm III}$ state the electron transfer goes in the opposite direction, resulting in increase of its oxidation state to U$^{\rm IV}$.

The U substitutions are additionally studied with the use of multi-center models, which can provide both more structural and electronic relaxation and also include charge-compensating vacancies.
For ``starting'' U$^{\rm VI}$ case, the decrease in oxidation state similar to that of Am$^{\rm VI}$ and Cm$^{\rm VI}$ in one-center clusters is observed in our calculations but in a different way.

Since the really synthesized YNbTiO$_6$ structures can not be considered as perfect (periodic) crystals because the Nb and Ti atoms are statistically distributed within them occupying the same Wyckoff positions, different Ti $\leftrightarrow$ Nb substitutions are studied and corresponding structural changes are estimated.
\end{abstract}%

\maketitle


%===================	
	\section{Introduction}
	\label{Intro}
	
The YNbTiO$_6$ crystal is a synthetic end-member of the euxenite-(Y) or polycrase-(Y) mineral, (Y,Ca,Ce,U,Th)(Ti,Nb,Ta)$_2$O$_6$, which is in turn a member of a wide euxenite group with general formula AB$_2$O$_6$ (B = Ti,Nb,Ta; A can be a variety of metals). These minerals are known to contain rare earth and actinide atoms as natural impurities. Despite metamictisation due to radioactive impurities, the euxenite-group minerals were found to be considerably more resistant to both oxidation and leaching of uranium atoms than minerals of a closely related betafite/pyrochlore group (A$_2$B$_2$O$_7$ and similar)\cite{Khanmiri:18_polycrase, Khanmiri:18_betafite} and, therefore, they are considered as promising matrices for long-term high-level waste (HLW) storage.

The other important area of application of YNbTiO$_6$ is conditioned by its luminescence properties. YNbTiO$_6$ has self-activated luminescence\cite{Pei-2018} while also can serve as a host for rare earth (RE) doped phosphors. Due to equality of the oxidation states and similarity of ionic radii, Y$^{3+}$$\rightarrow$RE$^{3+}$ substitution does not significantly distort the crystal and the complex luminescence mechanism can be tuned by controlling the selection of dopants and their concentrations\cite{Ma-2009, Shi-2011, Ma-2013, Yu-2015, Venugopal-2018}.

Both above mentioned applications are united by the $f-$element (lanthanide and actinide)  impurities, which creates a considerable challenge for computational methods to simulate properties of interest with the accuracy sufficiently high for the applications.
 Direct \abinitio periodic-structure calculations of the compounds containing $f-$elements with required accuracy are in general impossible to-date and various semi-empirical methods are used in practice  for periodic-structure compounds even with heavy $p-$elements as substitutions\cite{Boutinaud-2020}. Moreover, additional problems arise when there is a difference in oxidation state between the original and substitute atoms (i.e., the substitution changes the original charge).

Last, but not least, Nb and Ti atoms are distributed statistically  in the practically synthesized YNbTiO$_6$ ``crystals'', while the \abinitio periodic-structure calculation can be performed only for a perfect crystal. Therefore, any such calculation will inevitably be only a rough approximation to a real structure.

One of the most reasonable ways to overcome these problems is to apply the embedding potential theory for a crystal fragment studies which makes it possible to simulate effective crystal surroundings in a limited size cluster calculation and, therefore, allowing one to use much more advanced methods and computational parameters, than that in conventional periodic-structure calculations.
This approach allows one to simulate local perturbations including point defects, such as substitutions and rearrangement of statistically distributed atoms (for example, Nb and Ti in  YNbTiO$_6$). Recently developed compound-tunable embedding potential (CTEP) method 
\cite{Lomachuk:20_pccp, Maltsev:21_prb, Shakhova:22_CTPP-YbHal_n} was already applied to a similar CaNb$_2$O$_6$ crystal, which is an end member of the fersmite mineral of the same euxenite group\cite{Maltsev:21_prb}; the CTEP accuracy was estimated by comparing the electronic densities in the embedded clusters with that of the original crystal; the effects of the basis set increase were tested and two ways of simulating Ca$\rightarrow$U substitutions were compared.

In the present study the CTEP method is applied to YNbTiO$_6$ and CaNb$_2$O$_6$ crystals. Though the calcium niobate electronic structure was previously studied in \cite{Maltsev:21_prb}, it was the pilot application of CTEP that was aimed mainly on estimating the applicability of the method to niobate crystals, and only Ca$^{2+}$$\rightarrow$U$^{4+}$ and Ca$^{2+}$$\rightarrow$U$^{6+}$ substitutions were briefly studied there. At the present study, the actinide substitutions (A$\rightarrow$M$^{n+}$; A=Y$^{3+}$,Ca$^{2+}$; M = U, Np, Pu, Am, and Cm; n=3,4,5,6) are modelled for both crystals and the resulting structures are studied thoroughly.

%=====================
 \section{CTEP method}

The detailed description of the CTEP method is given in our previous papers \cite{Lomachuk:20_pccp} \cite{Maltsev:21_prb}  \cite{Shakhova:22_CTPP-YbHal_n}. The general idea of CTEP is to select some crystal fragment and simulate the influence of the crystal environment by the CTEP operator, which is presented in the form of a linear combination of specific short-range semilocal pseudopotentials (PPs, see \cite{Goedecker:92a,Titov:99, Mosyagin:06amin, Mosyagin:16_An, Mosyagin:17_Ln, Oleynichenko:23_LibGRPP} and references) for the atoms of nearest environment and the long-range Coulomb potentials (which acting only within the fragment is taken into account) from optimized fractional point charges centered on both nearest and some more distant atoms of the environment in general.
 
 The crystal fragment is not constrained to have any symmetry corresponding to the crystal space group, however, it is considered as a set of alternating anionic and cationic ``coordination spheres'' around one or more  cationic or anionic centers, correspondingly.
  In short, four steps are necessary to generate a CTEP for a crystal fragment of interest (cluster below):

(1) High-quality (as high as possible without computational problems like basis set linear dependence arising) periodic-structure DFT calculation of the perfect crystal with geometry optimization. In general, medium-core PPs~\cite{Shakhova:22_CTPP-YbHal_n} with corresponding basis sets are used for heavy cations (particularly, for $d,f-$elements) and anions, while all-electron basis sets can be used for light anions.

(2) Generation of large-core ``compound-tunable'' pseudopotentials (lc-CTPP) for all cations. Initial approximation for lc-CTPP is prepared as a large-core pseudopotential built for the effective state of the original atom in crystal, with this state being preliminary obtained by one of the population analysis methods (note that lc-CTPPs for non-equivalent (at different Wyckoff positions) atoms of the same type are generated independently). Next, lc-CTPPs are optimized by variation of a selected set of parameters (i.e.\ effective radius/exponent or coefficients of primitive gaussians) with the criterion of minimization of root mean square (RMS) value of energy gradients to with respect to coordinates of atomic nuclei for a crystal with the original PP replaced by the generated lc-CTPP. 
The basis sets for the lc-CTPPs are generated from the original basis sets for the corresponding atoms in crystal by tuning the core part in order to match the behavior of the lc-CTPP pseudofunctions.
When lc-CTPPs are optimized, they are applied for any embedded cluster built for the given compound.

(3) Building a cluster from several parts: (a) ``main cluster'' which is equal to or contains a crystal fragment of interest and must consist of one or more cationic centers and all their direct anionic neighbors; (b) nearest cationic environment (NCE) which contains the cations neighboring the main cluster; and (c) nearest anionic environment (NAE), which contains the anions neighboring the NCE except the main-cluster anions. Such generated environment is sufficient for our clusters, however, note, that for the main clusters with more complicated or oblong structures, the number of NCE and NAE spheres can be twice more.
The main-cluster atoms are initially treated by the same pseudopotentials and basis sets as in the solid-state calculation; NCE is represented by ``pseudoatoms'', modeled by lc-CTPPs and reduced basis sets combined with partial point charges; and NAE are represented by negative partial point charges only (however, in some cases simplified PPs can be added to the NCE pseudoatoms). All the atoms and pseudoatoms are located at the theoretically optimized lattice sites of corresponding atoms in the original crystal to provide correct reproducibility of the main-cluster electronic structure when generating CTEP in addition to the above mentioned constraints on the PPs and basis sets. However, in some cases arbitrary point charges outside the main cluster can also be in principle added to reproduce long-range electrostatic potential.

Initial distribution of the partial charges is obtained by solving a system of linear equations for the charge transfer:
	\begin{equation*}
\sum_{j}^{j \in \rm{neighbors}(i)}{\rm CT}_{i\to j}={\rm RedOx}_{i}\ ,
\end{equation*}
where ${\rm CT}_{i\to j}$ is estimated formal charge transfer from atom $i$ to $j$ (${\rm CT}_{j\to i}=-{\rm CT}_{i\to j}$), and ${\rm RedOx}_{i}$ is the oxidation state of the corresponding atom in the crystal. The system is usually an underdetermined one, so a minimum-norm solution is used.

When ${\rm CT}_{i\to j}$ are found, the partial charges are estimated as
\begin{equation*}
Q_{i \in \rm NCE} = \rm RedOx_{i}\ ,
\end{equation*}
and
\begin{equation*}
Q_{i \in \rm NAE} = \sum_{j}^{j \in \rm NCE}{\rm CT}_{i\to j}\ .
\end{equation*}

(4) ``Optimization'' of NCE and NAE point charges, made by variation of the charges within certain limits (usually, from zero to the oxidation number) with the criterion of minimization of RMS forces on the nuclei of the main cluster.

After these steps are successfully completed, the preparation of the embedding potential is finished. The resulting cluster with CTEP can be used in various ways. One can increase the precision of calculation either by applying one of advanced wave-function-based correlation methods\cite{oleynichenko2023compoundtunable} instead of DFT or by increasing basis set and/or using small-core PPs. Different atom-in-compound (AiC) properties on heavy atoms (see detailed discussion about it in papers\cite{Titov:14a, Lomachuk:13, Skripnikov:15b, Zaitsevskii:16a, Oleynichenko:18_AiC, Lomachuk:18en}) can be studied within the two-step approach (see details in papers \cite{Titov:85Dism, Kozlov:87, Dmitriev:92,Titov:05_IJQC}). Diversity of structural and electronic perturbations can be  considered in the main cluster, and point defects can be simulated, where one or multiple centers are replaced by the other atom or vacancy (note that the positions and charges of NCE and NAE are fixed for all perturbations and replacements).

%==============================
\section{Computational details}
	
The {\sc crystal} code \cite{CRYSTAL17} with unrestricted hybrid PBE0 functional \cite{Adamo:99} was used to carry out periodic electronic structure calculations of crystals.
A slightly modified {\sc so-dft} code from the {\sc nwc}hem package \cite{NWChem:10_CPC} was used for the unrestricted DFT PBE0 calculations of the clusters.

We used basis sets and pseudopotentials generated by our group~\cite{Mosyagin:17_Ln} for metal centers. For CaNb$_2$O$_6$, all parameters were same as in our previous work \cite{Maltsev:21_prb}. For YNbTiO$_6$, basis sets were built by a slightly modified approach, so that there are minor differences between basis sets of Nb in CaNb$_2$O$_6$ and YNbTiO$_6$. All PPs and basis sets used for solid state calculations and single-center CTEP clusters are presented in supplementary materials. 

%=================================	
\section{Results and discussions}

%----------------------------------- 
\subsection{Periodic structure calculation of YTiNbO$_6$}

\begin{figure}[h]
	\centering
	\includegraphics[width=0.75\linewidth]{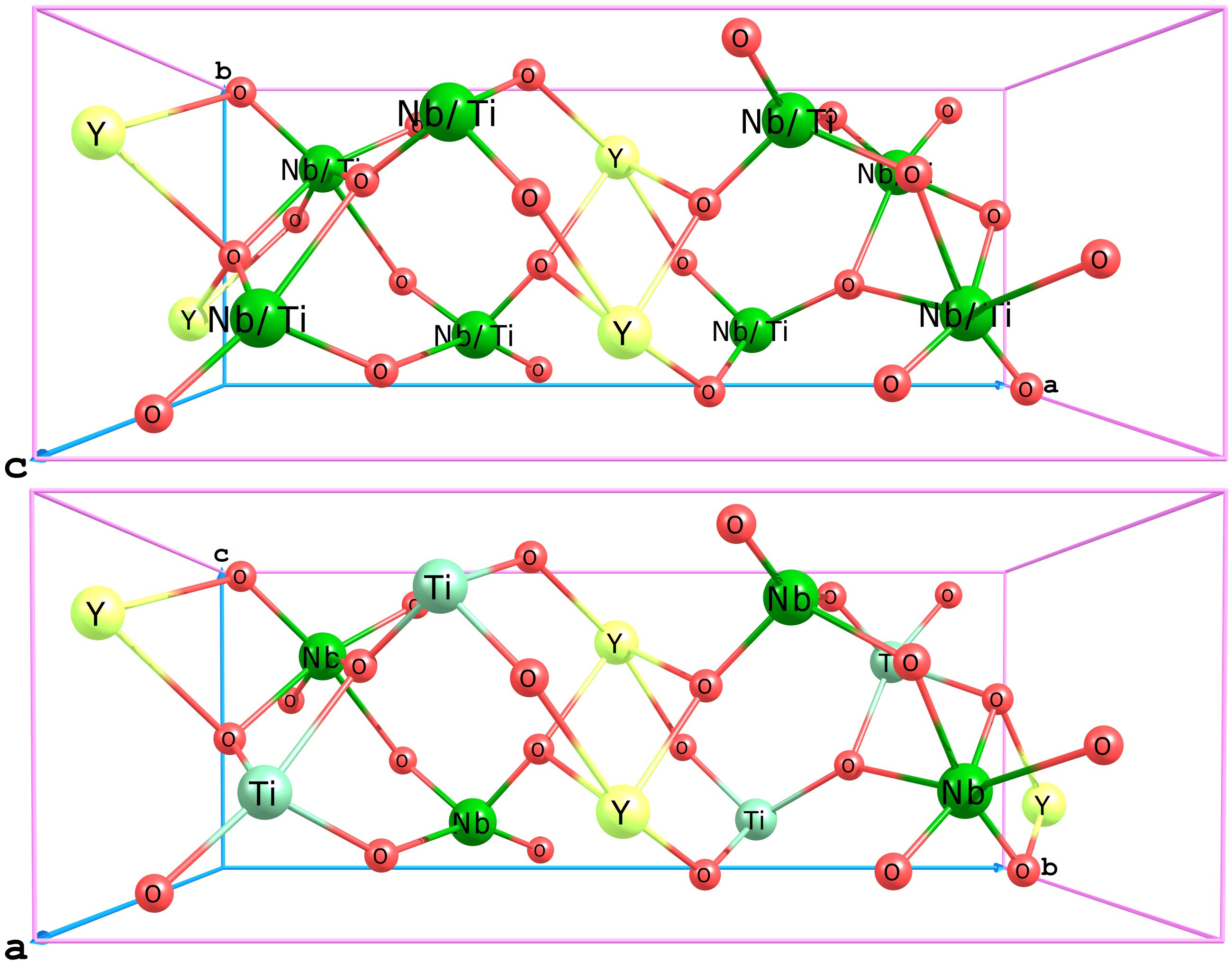}
	\caption{
  Statistically-averaged experimental (top) and calculated (bottom) conventional cells for YNbTiO$_6$.
	}
	\label{fig:polycrase_crystal}
\end{figure}

 In real crystals the Ti and Nb atoms in YTiNbO$_6$ are distributed statistically, occupying the same position, each with 50\% probability. Such structure can not be calculated directly, so a model crystal was constructed, in which the positions of Ti and Nb are split (Figure \ref{fig:polycrase_crystal}). The symmetry of the resulting cell was lowered from Pbcn (60) space group to P2$_1$/c (14) with only two of cell angles remaining right. Several variants of splitting were tested and the lowest by energy was selected.
	
The calculated and experimental parameters are compared in Table~\ref{table:polycrase_crystal}. Overall agreement is satisfactory, considering inevitable distortions resulting from splitting of the Ti and Nb positions and lowering the symmetry.

\begin{table}[h!]
	\caption{Experimental and calculated structure parameters for YNbTiO$_6$. Cell vector order differs due to change of the space group in the calculated structure.}
	\label{table:polycrase_crystal} 
	\begin{tabular}{lrr}
		\hline
		Parameter & Exp.\cite{Weitzel+1980+69+82} & Calc. \\  
		\hline
		a, \AA & 14.643 & 5.166 \\ 
		\hline 
		b, \AA & 5.553 & 14.649 \\ 
		\hline 
		c, \AA & 5.195 & 5.543 \\ 
		\hline 
  	  $\alpha$ & 90$^{\circ}$ & 90$^{\circ}$ \\ 
		\hline 
		$\beta$ & 90$^{\circ}$ & 89.45$^{\circ}$ \\ 
		\hline 
		$\gamma$ & 90$^{\circ}$ & 90$^{\circ}$ \\ 
		\hline 
		Y-O, \AA & 2.300--2.497 & 2.268--2.530 \\ 
		\hline 
		Ti-O, \AA & 1.722--2.408 & 1.825--2.241 \\ 
		\hline 
		Nb-O, \AA & 1.722--2.408 & 1.762--2.297 \\ 
		\hline 
	\end{tabular}
\end{table}

%----------------------------------------------------
\subsection{One-center clusters for perfect CaNb$_2$O$_6$ and YNbTiO$_6$ crystals}

\begin{figure}[h]
	\centering
	\includegraphics[width=0.95\linewidth]{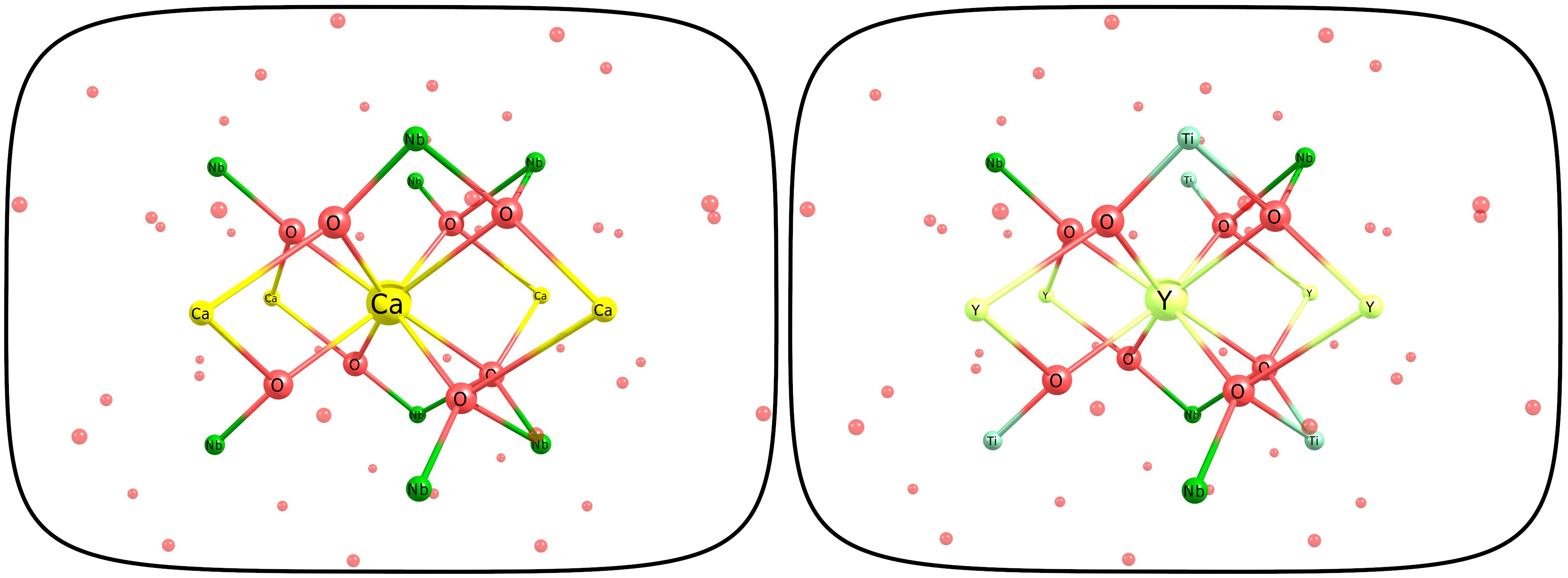}
	\caption{
		Ca-centered main cluster in CaNb$_2$O$_6$ (left) and Y-centered main cluster in YNbTiO$_6$ (right).
	}
	\label{fig:clusters_a}
\end{figure}

Crystals of the euxenite group have general formula AB$_2$X$_6$, where $f-$elements usually replace atom of A group. One-center clusters with AO$_8$@CTEP formula for CaNb$_2$O$_6$ (A=Ca) and YNbTiO$_6$ (A=Y) are presented on Figure \ref{fig:clusters_a}, the former being taken from our previous work\cite{Maltsev:21_prb}. Structures and partial charges for both clusters are presented in the supplementary materials.

Reproducing the perfect crystal geometry for Y-centered main cluster in YNbTiO$_6$ was found to be on the same level as the Ca-centered main cluster in CaNb$_2$O$_6$ (Table \ref{table:clusteropt_a}).
\begin{table}[h!]
	\caption{Forces on the atoms of the main 
		cluster and atomic displacements within the main cluster after its optimizations}
	\label{table:clusteropt_a} 
	\begin{tabular}{lcc}
		\hline
		Structure & RMS force (a.u.) & RMS displacement({{\AA}}) \\  
		\hline
		Ca$_c$ (CaNb$_2$O$_6$) & 2.6$\cdot$10$^{-5}$ & 3.4$\cdot$10$^{-4}$ \\ 
		\hline
		Y$_c$ (YNbTiO$_6$) & 1.9$\cdot$10$^{-5}$ & 1.9$\cdot$10$^{-4}$ \\ 
		\hline 
	\end{tabular}
\end{table}

%----------------------------------
\subsection{Basis set increase effects}

After the cluster is successfully built, it is possible to modify calculation parameters in order to achieve specific goals. One of the simplest modifications is the increase of the basis sets on one or multiple atoms and/or pseudoatoms, which allows one to answer two questions: (1) how stable is the main cluster geometry towards the variation of the basis sets, and (2) can the basis set increase lead to more accurate geometry (closer to the experimental one). However, the latter test is hampered by the fact that the embedding potential is built for non-experimental geometry, which means that CTEP rather mitigates only a part of deviations of the discussed theoretical model from the experiment.
Additionally, for YNbTiO$_6$ there is a specific inevitable source of geometry errors, arising from the lowering of the symmetry in order to build a perfect crystal.

\begin{figure}[h]
	\centering
	\includegraphics[width=0.95\linewidth]{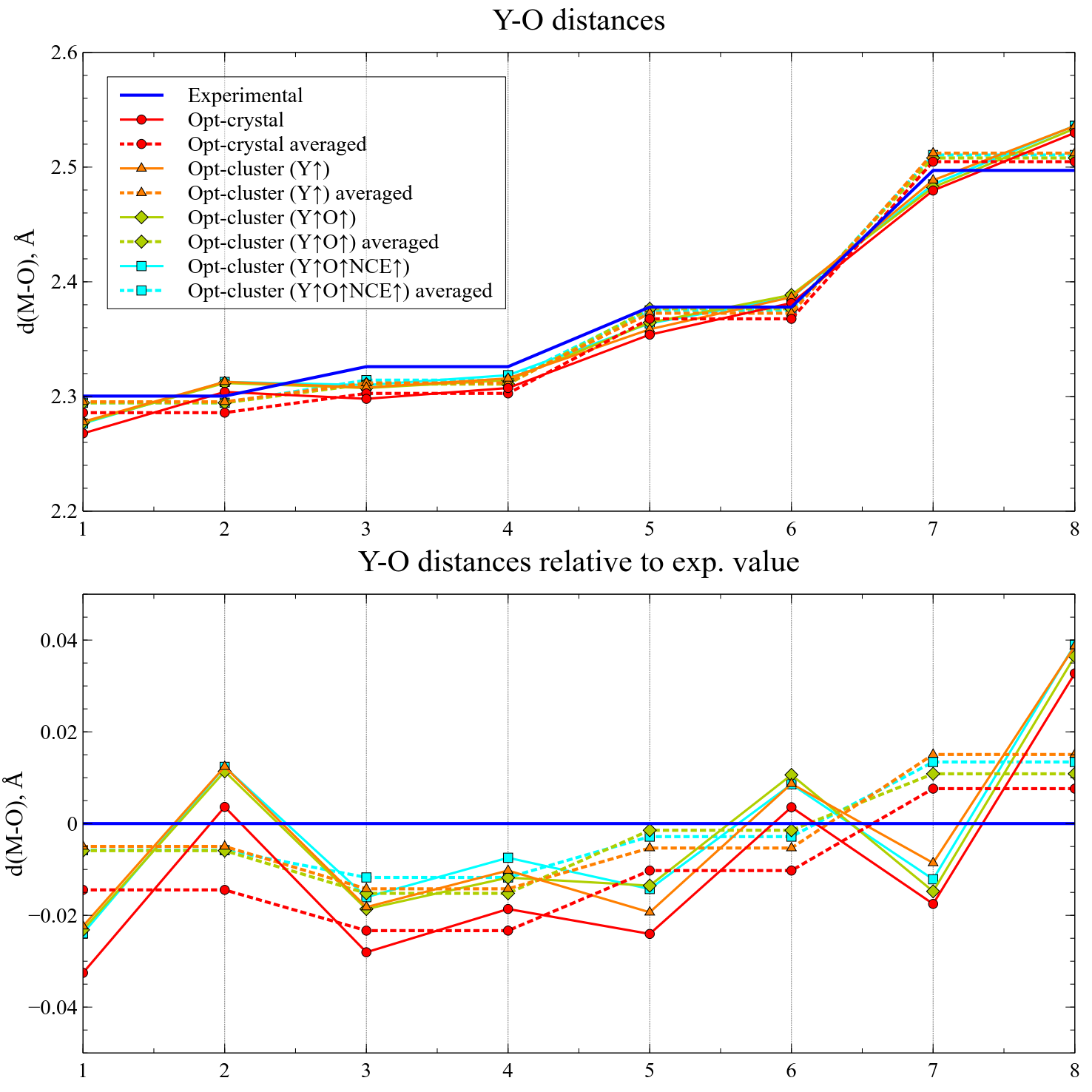}
	\caption{
		The Y-O distances in the one-center main cluster for different basis sets. Numbers on x axis represent index of each O neighbor of the central Y atom.  For each structure real (solid lines) and averaged (dashed lines) Y-O distances are shown. The averaging is made for pairs of O atoms located on equal distances from Y atom in the experimental structure. ``Opt-crystal'' corresponds to both periodic optimized structure and cluster with the same basis sets. Up arrows denote increased basis for the corresponding atoms.
	}
	\label{fig:polycrase_basis}
\end{figure}

Several basis set modifications were made: starting with the original parameters from the crystal calculations, basis sets were successively increased in the valence area for (1) the central Y atom, (2) its O neighbors, and (3) NCE pseudoatoms (see supplementary materials for all basis sets involved). The obtained results compared to the experimental structure are presented on Figure \ref{fig:polycrase_basis}. 
The Y-O distances in the experimental structure
are grouped into pairs due to local symmetry, so, for ease of comparison, each cluster corresponds
to two graphs: the real geometry (solid line), and distances averaged by pairs (dashed line). Additionally, to make it less cluttered, differences from the experimental values are shown on the bottom graph.

First tendency seen on the graph is that all calculations slightly underestimate the lower 6 Y-O distances and overestimate the distance to the 2 farthest O atoms. The latter is probably the result of the former effect: as the closest atoms become more closer, the remaining two are pushed out. The increase of Y basis set leads to almost equal increase of all eight Y-O distances. Additional increase of the oxygen basis sets leads, however, to additional increase of some of the first six Y-O distances and to decrease of the longest two Y-O distances, thus resulting in a structure closer to the experimental one. The subsequent increase of NCE basis sets leads to slight increase of the longest Y-O bonds and small variations of the shortest six ones. Thus, the overall closest structure to the experimental one was obtained when the basis sets on Y and O are increased, but the NCE pseudoatoms keep the original basis sets.

These tendencies can be partly explained by the assumption that the basis set increase indeed leads to a structure closer to the experimental one, however, as the embedding potential was built to keep the optimized structure, equal basis increase for both atoms and pseudoatoms prevents such change, which is especially significant for the two farthest O atoms, which are bound stronger to the NCE atoms than to the central Y. Last but not least, one should remember that realistic experimental structure is unachievable in any periodic structure calculations because of random distribution of Nb and Ti atoms.  Besides, the computational uncertainties also influence on the theoretical result.

%----------------------------------
\subsection{Actinide substitutions}

Both clusters which are simulating fragments of calcium niobate and perfect yttrium titano-niobate crystals were used for modeling A$\rightarrow$ M$^{n+}$ substitution (A = Ca$^{2+}$ for CaNb$_2$O$_6$ and Y$^{3+}$ for YNbTiO$_6$; M = U, Np, Pu, Am, Cm; n = 3, 4, 5, 6), with total of 40 actinide-substituted clusters being constructed. The structural relaxation (with NCE and NAE pseudoatoms being fixed) was performed and 
distribution of spin density was calculated for each substituted cluster.

\begin{figure}[h]
	\centering
	\includegraphics[width=0.95\linewidth]{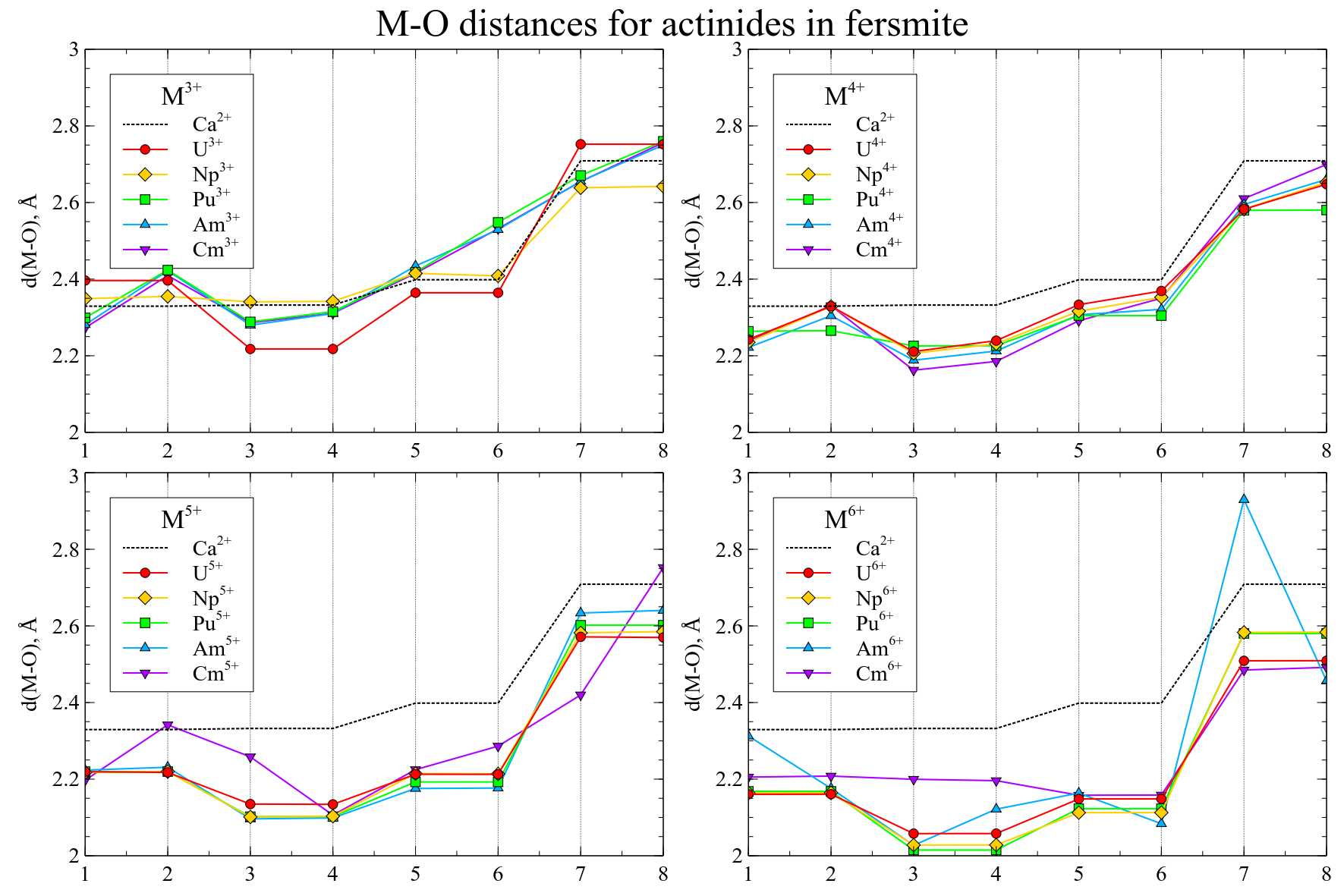}
	\caption{
		The M-O distances for actinide substitutions in CaNb$_2$O$_6$. Numbers on the $x$ axis represent indices of each O neighbor of the central atom. Dashed line corresponds to Ca-O distances in the perfect crystal.
	}
	\label{fig:structures_f}
\end{figure}

\begin{figure}[h]
	\centering
	\includegraphics[width=0.95\linewidth]{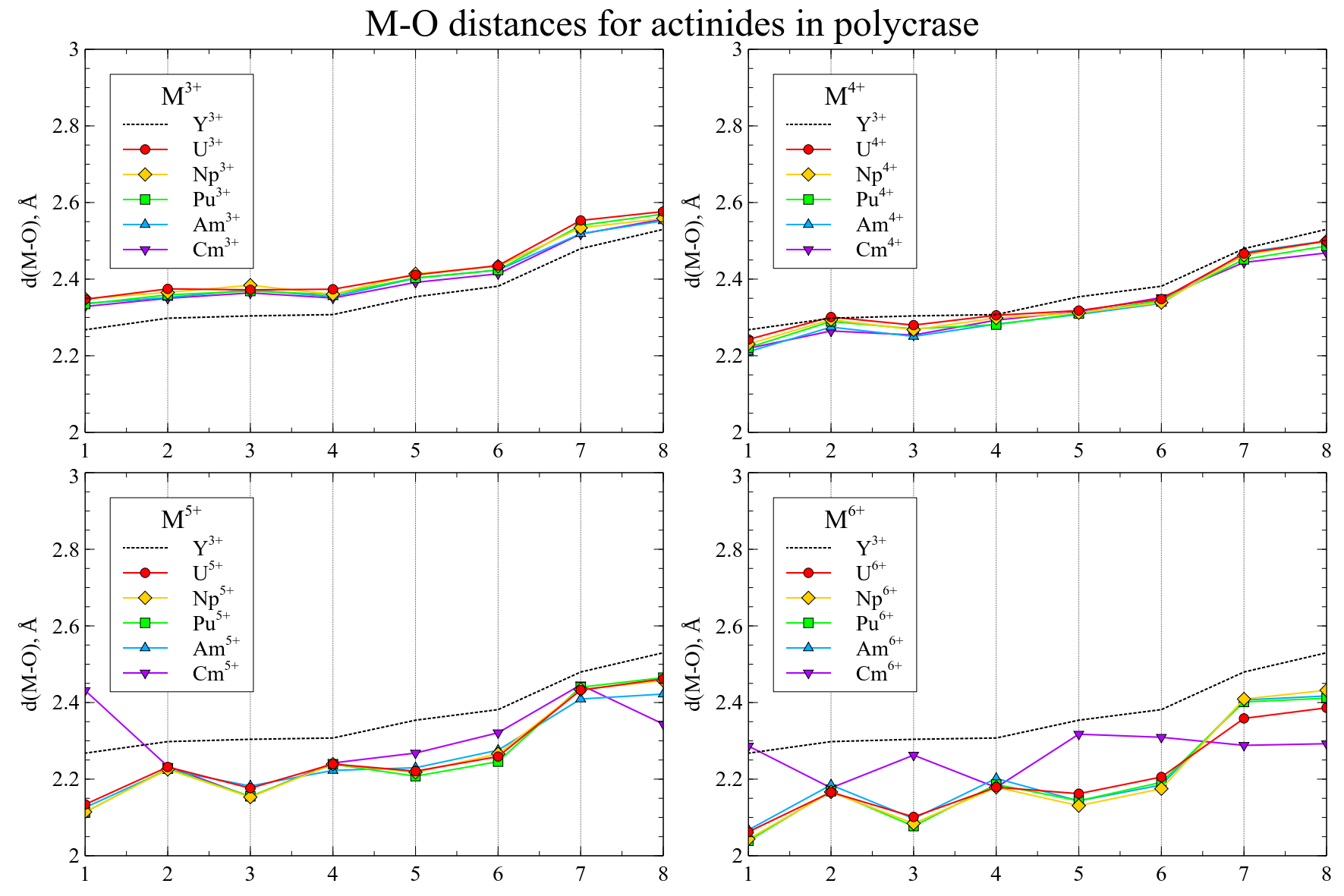}
	\caption{
		The M-O distances for actinide substitutions in YNbTiO$_6$. Numbers on the $x$ axis represent induces of each O neighbor of the central atom. Dashed line corresponds to Y-O distances in the perfect crystal.
	}
	\label{fig:structures_p}
\end{figure}

On Figures \ref{fig:structures_f} and \ref{fig:structures_p} the relaxed main cluster structures are presented as a set of M-O distances.

\begin{figure}[h]
	\centering
	\includegraphics[width=0.95\linewidth]{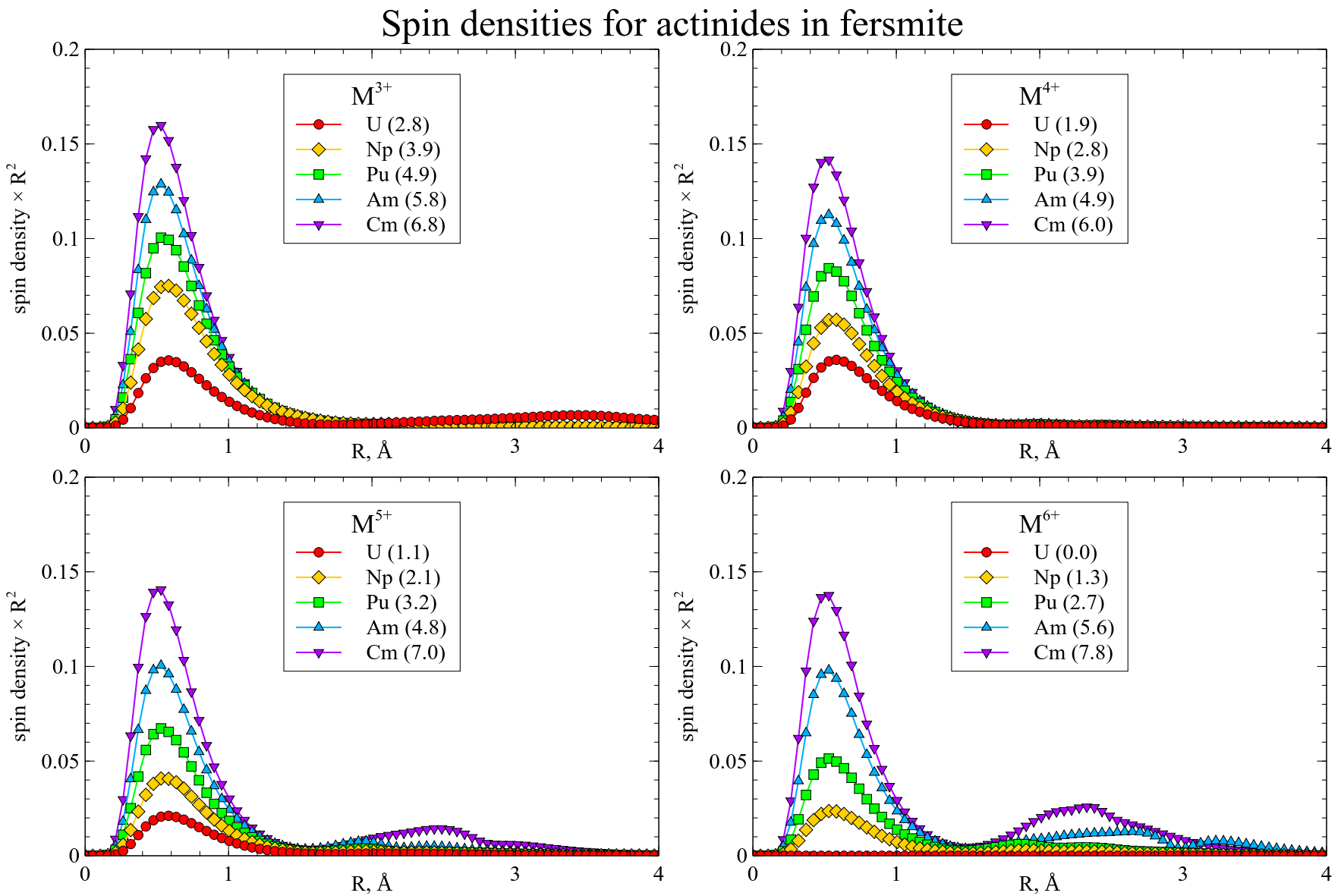}
	\caption{
		Integrated spin densities for actinide substitutions in CaNb$_2$O$_6$. Total values of spin are given in parentheses in the legend.
	}
	\label{fig:spindens_f}
\end{figure}

\begin{figure}[h]
	\centering
	\includegraphics[width=0.95\linewidth]{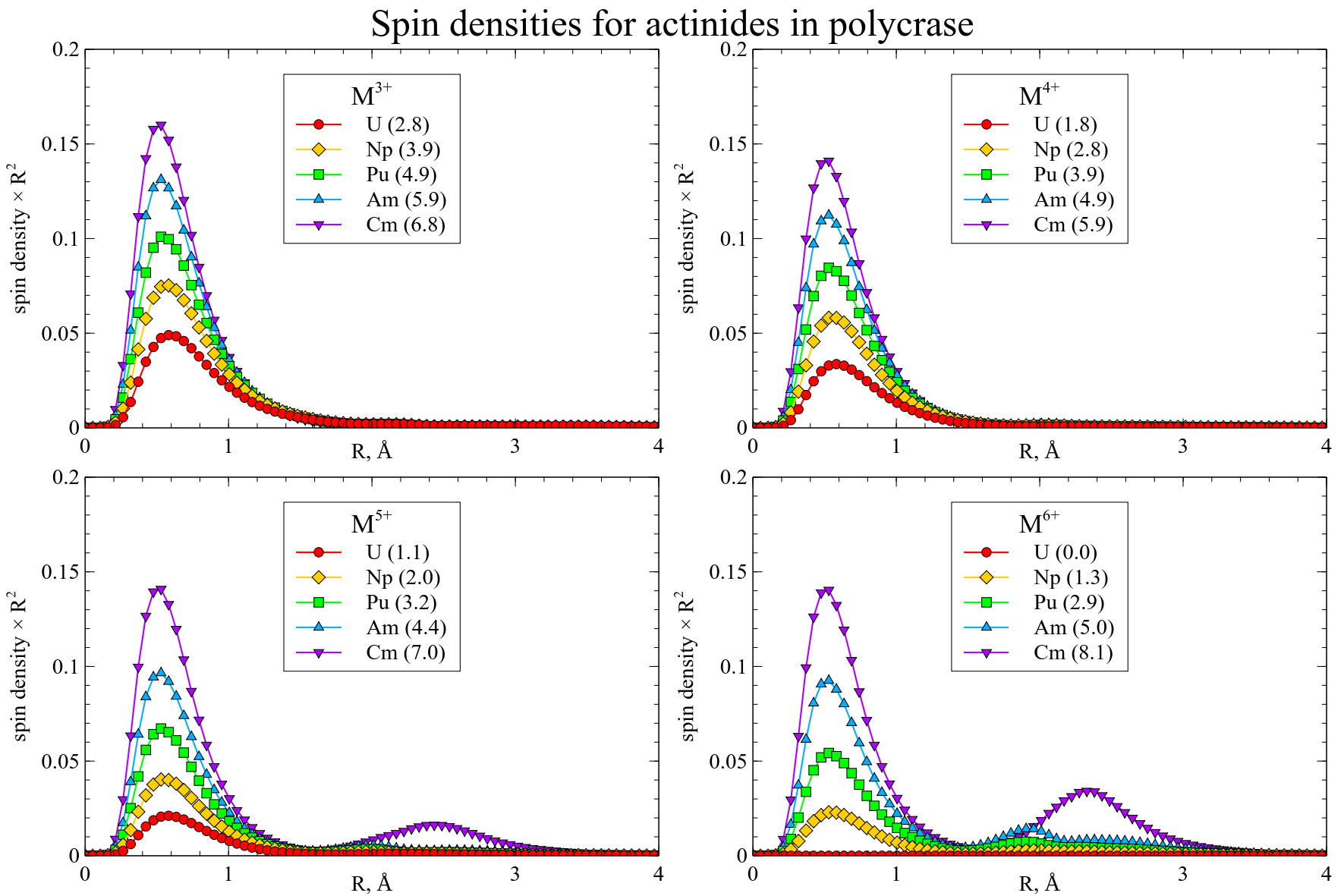}
	\caption{
		Integrated spin densities for actinide substitutions in YNbTiO$_6$. Total values of spin are given in parentheses in the legend.
	}
	\label{fig:spindens_p}
\end{figure}

The actinides in the same oxidation state, as a rule, yield similar structures. Thus one can suppose that such substitutions will exhibit similar physical and chemical properties and data for uranium can be extrapolated on the rest actinides under study. However, there are several apparent exceptions from this rule: U$^{3+}$ in CaNb$_2$O$_6$ and Am, Cm in high oxidation states in both structures. To investigate this deviations and obtain more detailed data, the spin densities were calculated for each cluster and integrated as function of distance to the central atom:
	\begin{equation*}
sd(r) = \frac{1}{4\pi}\oint d \Omega \left | \rho_{spin}(\vec{r}) \right |
\end{equation*}
The resulting graphs are shown on Figures \ref{fig:spindens_f} and \ref{fig:spindens_p}. The maxima of the spin density, associated with the radii of $f$-orbitals are almost equal for all clusters. For most of the clusters the spin density is localized on the actinide atom and the number of unpaired electrons increases with increase of the atomic number and decreases with increase of the oxidation state. 

However, there are several deviations. For U$^{3+}$ in CaNb$_2$O$_6$ spin density within U atomic radius is similar to that of U$^{4+}$ and there is a second wide maximum, which indicates that electron is transferred from central U to NCE of embedding potential. For Cm$^{6+}$, Cm$^{5+}$ and Am$^{+6}$ the electron is transferred in a reverse direction: from neighbor oxygen atoms to the actinide, so that maximum obtained oxidation states for Cm and Am are +4 and +5, correspondingly.

While such results can be explained by inaccuracy of the applied model, the other interpretation is that the obtained results can serve as an indication that the mentioned oxidation states of the actinides as substituents cannot exist in the considered fragments of niobate matrices in practice.

%---------------------------------
\subsection{X-ray line chemical shifts}

Chemical shifts of the lines of X-ray emission spectra (chemshifts) for the actinides were calculated
for all substitutions under study. 
Our results are collected in Tables~\ref{table:shifts_f} and \ref{table:shifts_p}.

\begin{table}[]
	\caption{X-ray line chemical shifts (meV) for actinides in CaNb$_2$O$_6$ relative to free M$^{4+}$ ion}
	\label{table:shifts_f} 
	\begin{tabular}{l|rrrr}
		& K$\alpha_2$    & K$\alpha_1$    & L$\beta_4$   & L$\beta_3$    \\
		\hline
		U$^{[3+]}$  & 19.7   & 30.2   & 63.4   & 103.7  \\
		U$^{4+}$  & 28.7   & 41.6   & 77.8   & 120.0  \\
		U$^{5+}$  & 171.0  & 226.7  & 163.8  & 255.5  \\
		U$^{6+}$  & 252.1  & 332.5  & 207.9  & 352.4  \\
		\hline
		Np$^{3+}$ & -180.0 & -229.9 & -162.7 & -118.4 \\
		Np$^{4+}$ & -5.3   & -7.1   & 31.3   & 71.3   \\
		Np$^{5+}$ & 132.0  & 169.3  & 141.2  & 222.3  \\
		Np$^{6+}$ & 217.7  & 278.1  & 202.4  & 314.0  \\
		\hline
		Pu$^{3+}$ & -185.4 & -243.3 & -77.0  & -149.9 \\
		Pu$^{4+}$ & -3.8   & -9.8   & 124.4  & 44.1   \\
		Pu$^{5+}$ & 109.9  & 136.9  & 221.2  & 162.7  \\
		Pu$^{6+}$ & 175.2  & 219.3  & 278.1  & 225.0  \\
		\hline
		Am$^{3+}$ & -176.2 & -230.2 & -85.2  & -100.7 \\
		Am$^{4+}$ & -13.8  & -21.8  & 85.2   & 46.8   \\
		Am$^{5+}$ & 64.8   & 78.9   & 142.0  & 119.2  \\
		Am$^{[6+]}$ & 70.5   & 85.7   & 149.9  & 127.9  \\
		\hline
		Cm$^{3+}$ & -197.3 & -254.4 & -187.8 & -120.3 \\
		Cm$^{4+}$ & -38.9  & -50.9  & 24.8   & 24.2   \\
		Cm$^{[5+]}$ & -22.0  & -27.8  & 29.7   & 47.3   \\
		Cm$^{[6+]}$ & 2.8    & 3.8    & 63.1   & 71.3   \\
		\hline
	\end{tabular}
\end{table}

\begin{table}[]
	\caption{X-ray line chemical shifts (meV) for actinides in YNbTiO$_6$ relative to free M$^{4+}$ ion}
	\label{table:shifts_p} 
	\begin{tabular}{l|rrrr}
		& K$\alpha_2$    & K$\alpha_1$    & L$\beta_4$   & L$\beta_3$    \\
		\hline
		U$^{3+}$  & -123.0 & -157.6 & -38.4  & -7.9   \\
		U$^{4+}$  & 16.7   & 22.6   & 84.1   & 126.3  \\
		U$^{5+}$  & 141.9  & 185.6  & 170.6  & 260.7  \\
		U$^{6+}$  & 223.6  & 292.5  & 212.8  & 356.5  \\
		\hline
		Np$^{3+}$ & -181.4 & -231.6 & -163.3 & -120.8 \\
		Np$^{4+}$ & -0.9   & -1.4   & 34.8   & 73.7   \\
		Np$^{5+}$ & 137.6  & 176.6  & 145.0  & 223.9  \\
		Np$^{6+}$ & 222.3  & 284.4  & 204.6  & 311.8  \\
		\hline
		Pu$^{3+}$ & -185.4 & -242.7 & -81.6  & -148.6 \\
		Pu$^{4+}$ & -3.1   & -8.7   & 125.4  & 42.4   \\
		Pu$^{5+}$ & 115.7  & 144.5  & 229.4  & 163.8  \\
		Pu$^{6+}$ & 174.7  & 219.3  & 269.4  & 221.5  \\
		\hline
		Am$^{3+}$ & -173.1 & -223.7 & -118.6 & -86.3  \\
		Am$^{4+}$ & -9.1   & -15.5  & 88.4   & 50.3   \\
		Am$^{5+}$ & 82.7   & 102.0  & 173.6  & 127.3  \\
		Am$^{[6+]}$ & 101.9  & 124.9  & 173.9  & 144.5  \\
		\hline
		Cm$^{3+}$ & -194.6 & -250.3 & -185.6 & -117.3 \\
		Cm$^{4+}$ & -25.7  & -32.9  & 28.8   & 36.5   \\
		Cm$^{[5+]}$ & -22.6  & -28.6  & 28.3   & 43.3   \\
		Cm$^{[6+]}$ & -10.5  & -12.2  & 35.6   & 59.6   \\
		\hline
	\end{tabular}
\end{table}

In general, the values of chemshifts change monotonously with increase of oxidation state. Also, the values support the observation that high oxidation state of Am, Cm in both clusters and U$^{3+}$ state in CaNb$_2$O$_6$ were not attained.

%------------------------------------------------
\subsection{3-center clusters for perfect crystals}

Single-center clusters, while being easier to calculate, have inevitable limitations when a significant perturbation is introduced to the main cluster, such as replacement of the center atom by an element with very different chemical properties  and/or with different oxidation state/ionicity and atomic angular momenta for a given oxidation state. In such cases an extended cluster model should be rather build. It can simulate not only relaxation in a larger area, but also additional defects such as charge-compensating vacancy. 

\begin{figure}[h]
	\centering
	\includegraphics[width=0.95\linewidth]{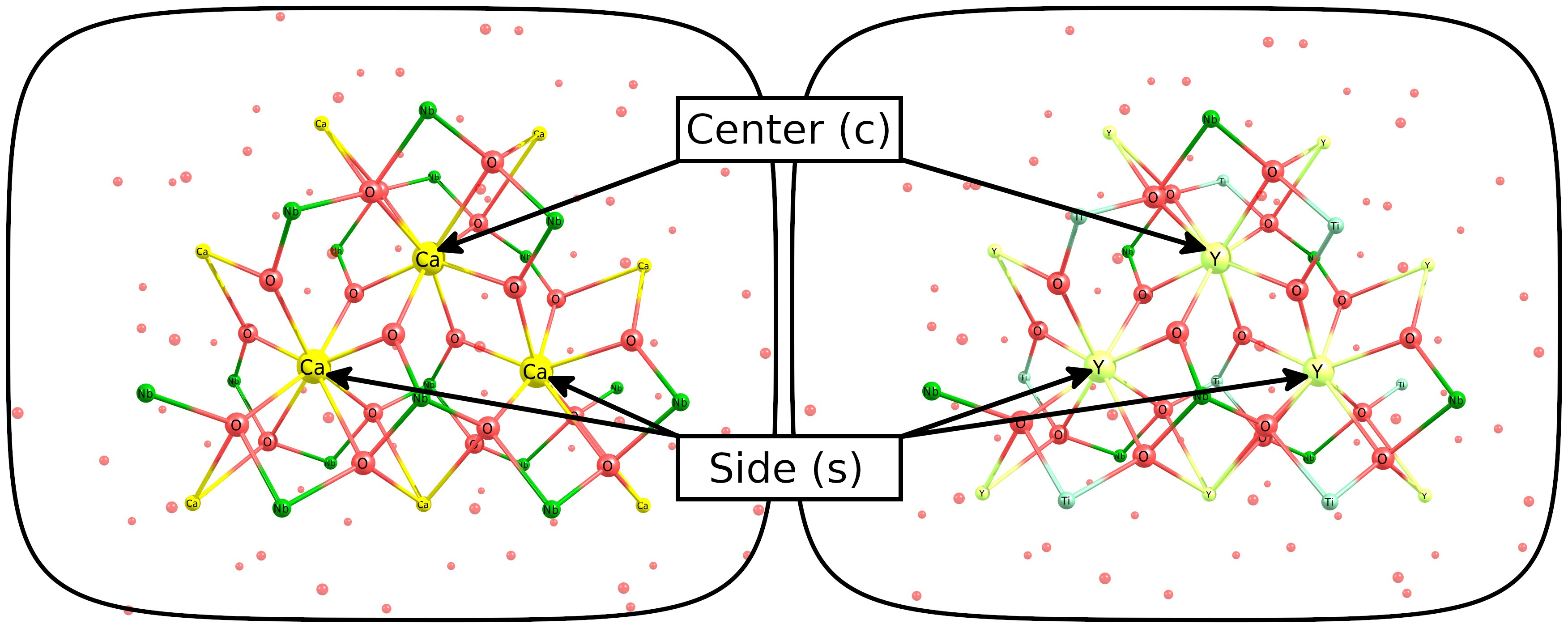}
	\caption{
		3-center clusters for Ca in CaNb$_2$O$_6$ (left) and Y in YNbTiO$_6$ (right). ``Center'' and ``side'' positions are denoted.
	}
	\label{fig:clusters_a3}
\end{figure}

For a more detailed study of actinides in niobates, 3-center cluster was constructed for both crystal systems (Figure \ref{fig:clusters_a3}). Clusters have C$_2$ symmetry, which is however slightly distorted for YNbTiO$_6$ due to lowering the original crystal symmetry. For a convenience, calcium centers are denoted as ``Center'' and ``Side''. 

The remaining RMS forces on the main clusters after their geometry optimization were about two orders of magnitude larger than those for the single-center model. While such values are still on par with common DFT errors, we managed to improve the results by several times by introducing and optimizing the lc-CTPPs on NAE pseudoatoms (Table \ref{table:clusteropt_a3}).

\begin{table}[h!]
	\caption{Forces on the atoms of the main 
		cluster.}
	\label{table:clusteropt_a3} 
	\begin{tabular}{lc}
		\hline
		Structure & RMS force (a.u.) \\  
		\hline
		3c-Ca$_3$ (CaNb$_2$O$_6$); no PP at NAE & 5.7$\cdot$10$^{-3}$  \\  
		\hline
		3c-Ca$_3$ (CaNb$_2$O$_6$); PP at NAE & 1.6$\cdot$10$^{-3}$  \\ 
		\hline
		3c-Y$_3$ (YNbTiO$_6$); no PP at NAE & 5.8$\cdot$10$^{-3}$ \\ 
		\hline
		3c-Y$_3$ (YNbTiO$_6$); PP at NAE & 1.7$\cdot$10$^{-3}$ \\ 
		\hline 
	\end{tabular}
\end{table}

%----------------------------------------
\subsection{Uranium in 3-center clusters}

\begin{figure}[h]
	\centering
	\includegraphics[width=0.95\linewidth]{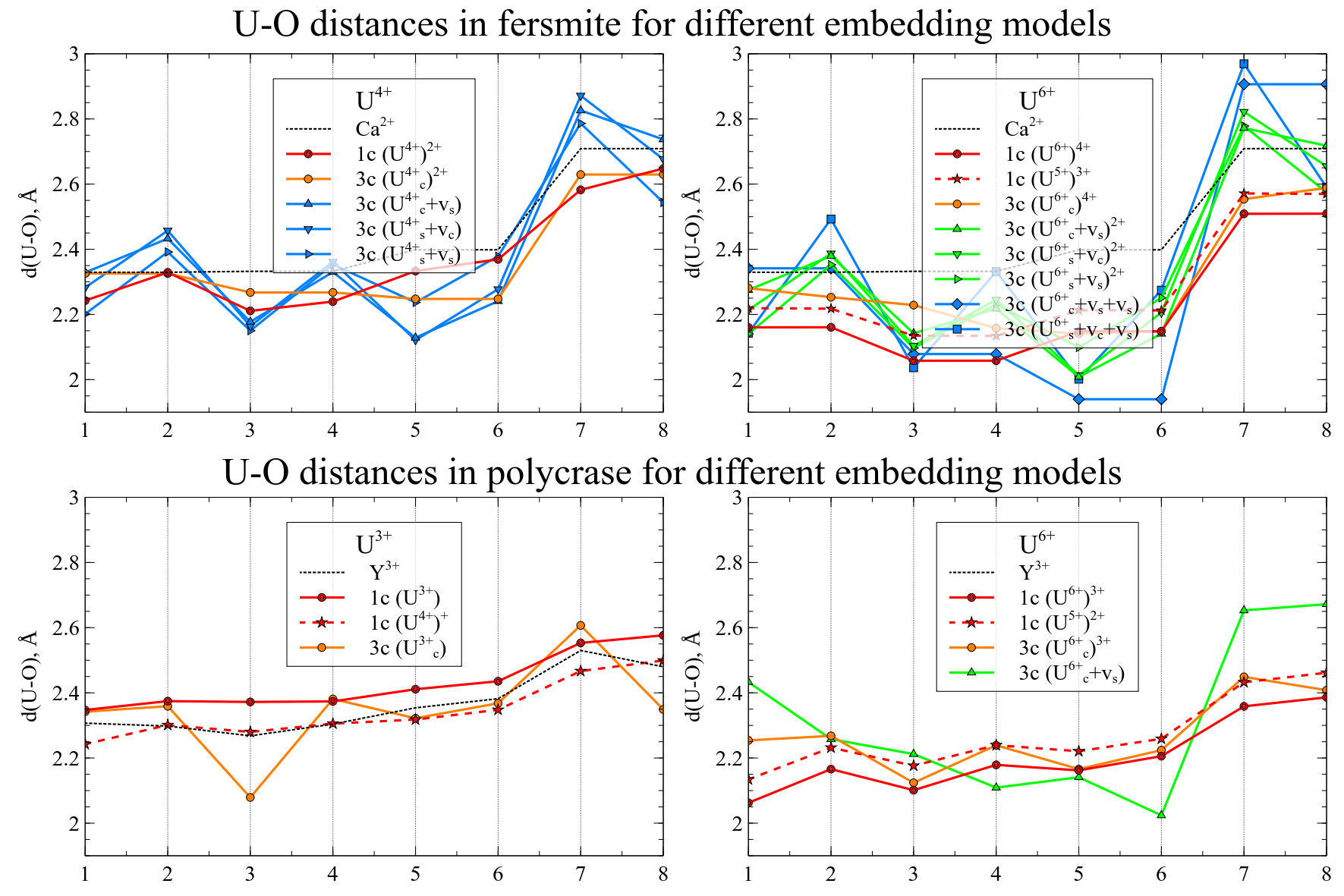}
	\caption{
		U-O distances for U substitutions in 3-center clusters for CaNb$_2$O$_6$ and YNbTiO$_6$. Numbers on x axis represent index of each O neighbor of the central atom. Dashed line corresponds to Ca-O and Y-O distances in perfect crystals. Red lines (solid and dashed) represent U-O distances in a one-center cluster substitution model.
	}
	\label{fig:structures_m}
\end{figure}

The 3-center clusters were used to simulate Ca$\rightarrow$U and Y$\rightarrow$U substitutions in CaNb$_2$O$_6$ and YNbTiO$_6$, correspondingly. From all previously studied oxidation states only those were chosen which allowed one to build a neutral embedded cluster with one uranium center and one or more compensating vacancies. In total, four substitutions were modeled: Ca$^{2+}$$\rightarrow$U$^{4+}$ (which can be compensated by a single Ca$^{2+}$ vacancy), Ca$^{2+}$$\rightarrow$U$^{6+}$ (which can be compensated by two Ca$^{2+}$ vacancies), Y$^{3+}$$\rightarrow$U$^{3+}$ (charge compensating is not required), and Y$^{3+}$$\rightarrow$U$^{6+}$ (which can be compensated by a single Y$^{3+}$ vacancy). Additionally, in all cases except Y$^{3+}$$\rightarrow$U$^{3+}$, the charged embedded clusters without vacancies (and with one vacancy for Ca$^{2+}$$\rightarrow$U$^{6+}$) were also built. Multiple arrangements of U and vacancies were modeled for CaNb$_2$O$_6$, but not for YNbTiO$_6$. The U-O distances in relaxed main clusters are presented on Figure \ref{fig:structures_m}. For comparison, the distances for the perfect crystal and U substitutions in single-center cluster model are also given.

Several conclusions can be made from the analysis of the resulting structures. For Ca$^{2+}$$\rightarrow$U$^{4+}$ substitution, the minimal (single) center model and charged 3-center model without vacancies yield a very similar structure, which means that the minimal model is  enough in practice to simulate an isolated substitution. As expected, the neighboring vacancy distort the structure, however all 3 arrangements of U atom and vacancy result in very similar structures.
In general, two cases of isolated substitution and substitution compensated by a vacancy must be considered when studying such substitutions and our model allows one to simulate both cases.

For Ca$^{2+}$$\rightarrow$U$^{6+}$ substitution, a charged cluster without vacancies yielded a structure which is closer to that of a one-center cluster with Ca$^{2+}$$\rightarrow$U$^{5+}$ substitution than that with Ca$^{2+}$$\rightarrow$U$^{6+}$. As in Ca$^{2+}$$\rightarrow$U$^{4+}$ substitution, 3 arrangements of a cluster with a single vacancy resulted in a similar structures, while two arrangements of the neutral cluster with two vacancies yielded two considerably different structures.

For the Y$^{3+}$$\rightarrow$U$^{3+}$ substitution, the resulting structure becomes closer to U$^{4+}$ one-center cluster, than the U$^{3+}$ one. For the Y$^{3+}$$\rightarrow$U$^{6+}$ substitution, the cluster without vacancies yields the structure intermediate between the U$^{5+}$ and U$^{6+}$ one-center clusters.

\begin{figure}[h]
	\centering
	\includegraphics[width=0.95\linewidth]{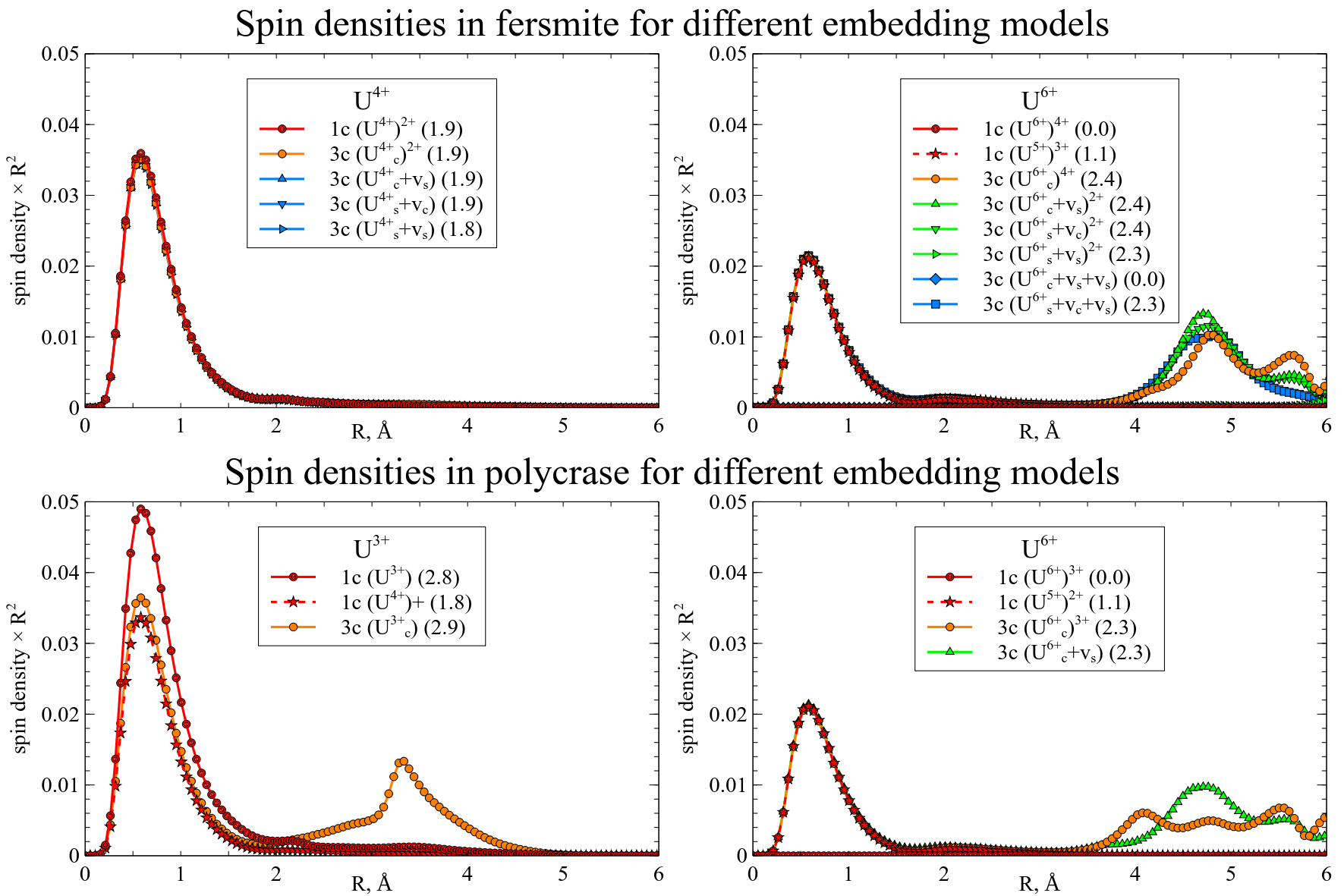}
	\caption{
		Integrated spin densities for U substitutions in 3-center clusters in CaNb$_2$O$_6$ and YNbTiO$_6$. Total values of spin are given in parentheses in the legend.
	}
	\label{fig:spindens_m}
\end{figure}

For a more detailed analysis, spin densities were calculated and integrated for each cluster (Figure \ref{fig:spindens_m}).

For Ca$^{2+}$$\rightarrow$U$^{4+}$ substitution all spin densities are similar, which means that U$^{4+}$ oxidation state is achieved in every cluster. However, for Ca$^{2+}$$\rightarrow$U$^{6+}$ substitution only neutral cluster with two vacancies yield the same spinless state as the single-center cluster. For all other cases the spin density in vicinity of the U atom is similar to that of U$^{5+}$ single-center cluster and there is another spin density peak at about 9 a.u. This indicates that the electron transfer to uranium takes place. For YNbTiO$_6$ both substitutions were found to undergo electron transfer: spin density of Y$^{3+}$$\rightarrow$U$^{3+}$ is close to that of Y$^{3+}$$\rightarrow$U$^{4+}$ in the minimal model, and spin density of Y$^{3+}$$\rightarrow$U$^{6+}$ is close to that of Y$^{3+}$$\rightarrow$U$^{5+}$ in the minimal model and the other other spin density peak is present in all cases.

\begin{figure}[h]
	\centering
	\includegraphics[width=0.95\linewidth]{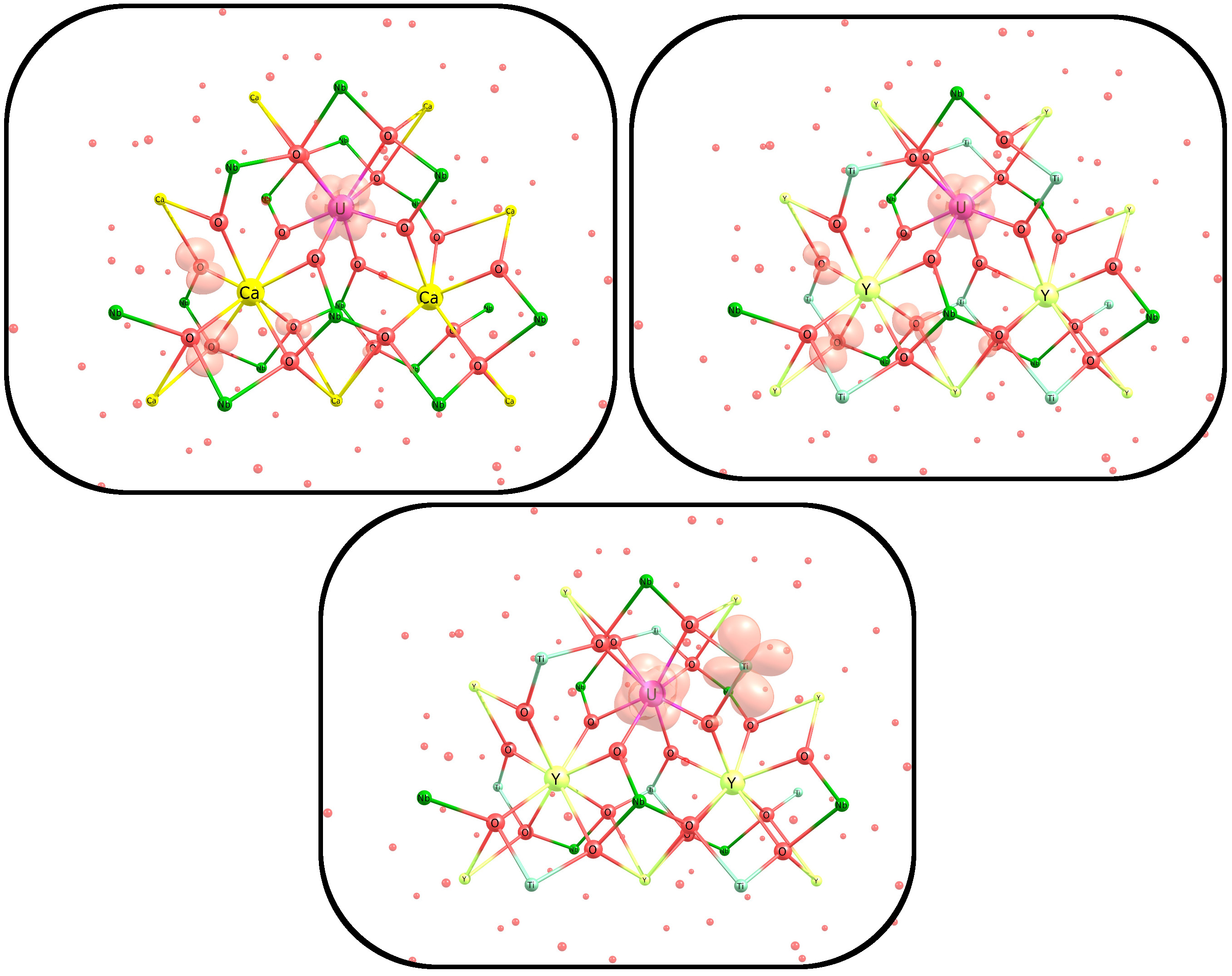}
	\caption{
		Spin densities for Ca$^{2+}$$\rightarrow$U$^{6+}$ substitution in CaNb$_2$O$_6$ and Y$^{3+}$$\rightarrow$U$^{6+}$ substitution in YNbTiO$_6$ (top) and Y$^{3+}$$\rightarrow$U$^{3+}$ substitution in YNbTiO$_6$ (bottom) for charged embedded clusters without vacancies.
	}
	\label{fig:spindens3_3d}
\end{figure}

To investigate the nature of these electronic transfers, 3d images of the spin density were made (Figure \ref{fig:spindens3_3d}).

\begin{figure}[h]
	\centering
	\includegraphics[width=0.95\linewidth]{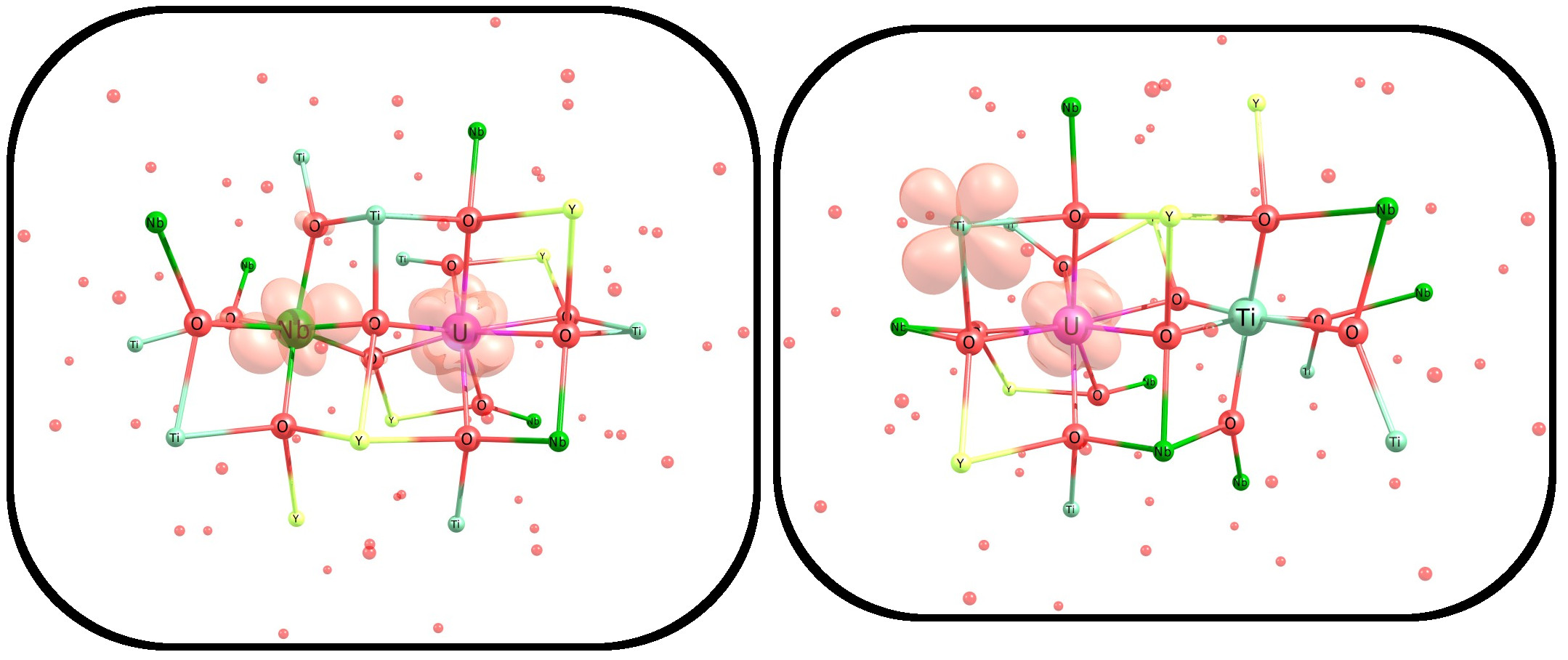}
	\caption{
		Spin densities for Y$^{3+}$$\rightarrow$U$^{3+}$ substitution in YNbTiO$_6$ for clusters with Y,Nb and Y,Ti centers.
	}
	\label{fig:spindens2_3d}
\end{figure}

                One can see that for U$^{6+}$ in CaNb$_2$O$_6$ and YNbTiO$_6$ O$\rightarrow$U the electron transfer takes place, however not from direct neighboring oxygen atoms, but instead from neighbors of the other center. For U$^{3+}$ there is electron transfer in opposite direction: from U to pseudoatom of the cationic layer in CTEP model. The latter rises a question: is such electron transfer to pseudoatom an indication of possible transfer to a real atom in crystal, or is it an error of CTEP model? To estimate the ability of electron transfer to real Ti and Nb atoms, two 2-center embedded clusters were built: with Y,Nb and Y,Ti centers. In each cluster Y$^{3+}$$\rightarrow$U$^{3+}$ were simulated with subsequent relaxation of structure, and images of spin densities were obtained (Figure \ref{fig:spindens2_3d}).

For U,Ti cluster, the U$\rightarrow$Ti electron transfer was not observed, the same transfer to pseudoatom as in 3-center clusters was found instead. However, for U,Nb cluster U$\rightarrow$Nb transfer occurred. The latter allows one to make a conclusion that U$^{3+}$ is a strong electron donor which cannot exist in a niobate matrices without being oxidized. And while it was found that the Ti pseudoatom is indeed a stronger electron acceptor than a ``normal'' Ti atom, it is a weaker acceptor than a ``normal'' Nb atom, so that the difference is not significant enough to invalidate the CTEP model.

%------------------------------
\subsection{Nb and Ti clusters}

\begin{figure}[h]
	\centering
	\includegraphics[width=0.95\linewidth]{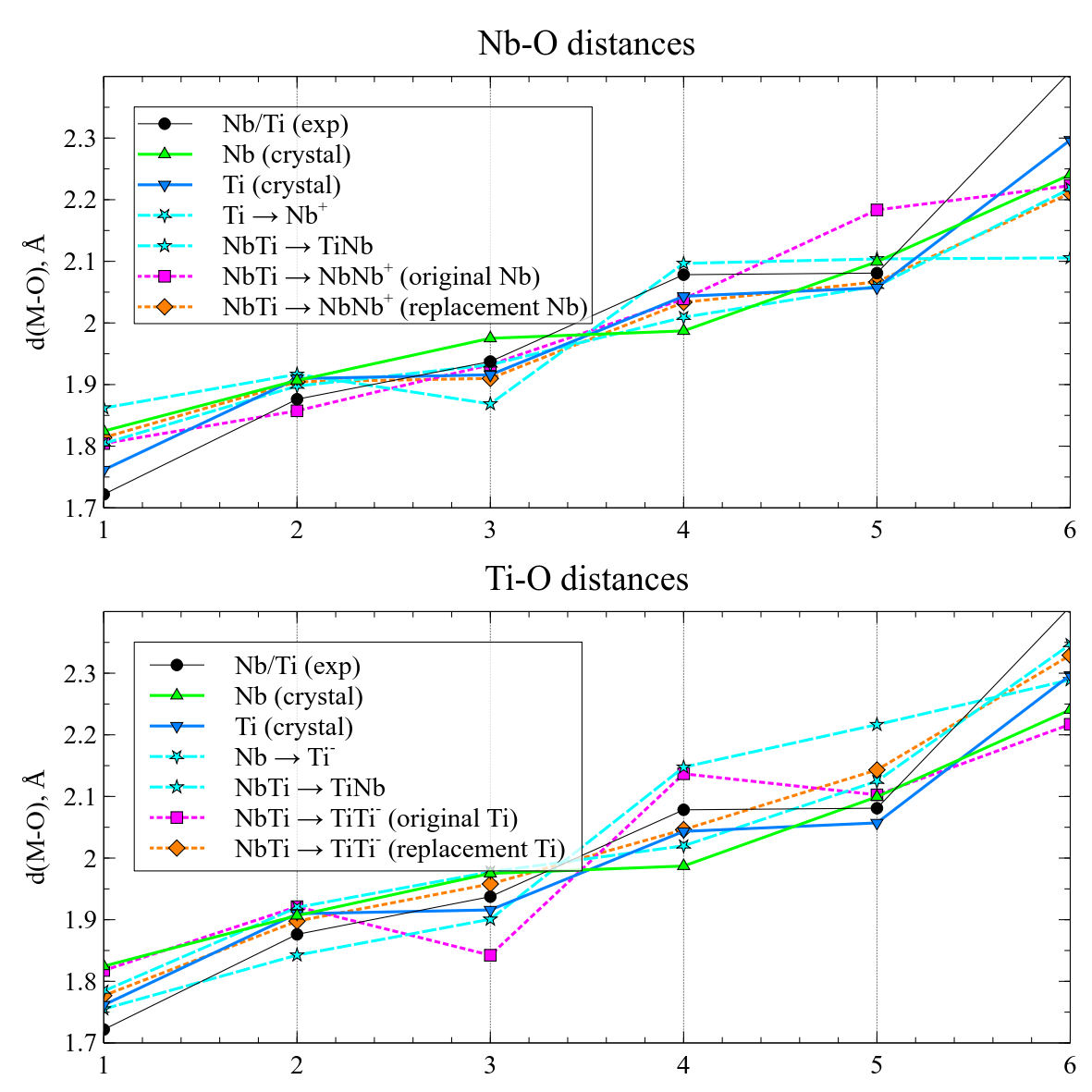}
	\caption{
		Structures of Nb/Ti substitutions in YNbTiO$_6$. Experimental data are given for comparison.
	}
	\label{fig:structures_nbti}
\end{figure}

As it was mentioned, while we use perfect crystals for our calculations, in real YNbTiO$_6$ Nb and Ti atoms are arranged randomly. The embedding potential method allows one to estimate the short-range effects of such rearrangement by modeling it as one or multiple substitutions in a corresponding cluster. In the present study, three clusters were built: minimal single-center clusters NbO$_6$@CTEP and TiO$_6$@CTEP and double-center NbTiO$_{10}$@CTEP cluster, and total of 5 substitutions were modeled with consequent relaxation of geometry:
\begin{enumerate}
\item Nb@CTEP $\rightarrow$ Ti@CTEP$^-$
\item Ti@CTEP $\rightarrow$ Nb@CTEP$^+$
\item NbTi@CTEP $\rightarrow$ Nb$_2$@CTEP$^+$
\item NbTi@CTEP $\rightarrow$ Ti$_2$@CTEP$^-$
\item NbTi@CTEP $\rightarrow$ TiNb@CTEP swap
\end{enumerate}
The results are displayed on Figure \ref{fig:structures_nbti}.

As it follows from the graph, the resulting structures exhibit considerable variability within certain range, so an assumption can be made that in the real polycrase crystal (with Nb and Ti randomly arranged) the structure of Ti and Nb surroundings exhibit variability within comparable range. While the CTEP method at the current stage cannot precisely describe such a structure, as the embedding potential still models the fixed perfect crystal environment, it definitely allows one to make a halfway step from the perfect crystal model to the real crystal without increase in the computation cost.

%~
%
%
%~

%====================
\section{Conclusions}
	
The CTEP method is applied to study actinide substitutions in two niobate crystals, CaNb$_2$O$_6$ and YNbTiO$_6$. Two one-center clusters are built (with Ca and Y centers, correspondingly), and 20 substitutions (five actinides, each in four different oxidation states) were made for each cluster. Geometry relaxation is performed for each resulting structure, and electronic properties are analyzed by evaluating the spin density distribution and calculating the X-ray line emission spectra chemical shifts. In general, the studied embedded clusters with actinides having the same oxidation state are found to yield similar local structure distortions that indicates the similarity of behavior of these elements in the studied niobate matrices.

For some cases, however, electron transfer was found to take place. Am and Cm in high oxidation ``starting'' states accept electrons from the neighbor O atoms, while ``starting'' U$^{\rm III}$ donates an electron to the second-order (cationic) neighbors. Both results indicate that the mentioned elements cannot exists in these oxidation state in the studied niobate matrices and undergo reduction or oxidation, respectively.

The U substitutions are additionally studied with the use of multi-center models, which can provide both more structural and electronic relaxation and also include charge-compensating vacancies. For ``starting'' U$^{\rm III}$ state, the electron transfer to the Nb cation is confirmed. For ``starting'' U$^{\rm VI}$ state, the reduction similar to that of Am$^{\rm VI}$ and Cm$^{\rm VI}$ in one-center clusters was found, except that the electron is donated by oxygen atoms from third-order coordination sphere and not by the nearest ones. This result seems to contradict the experimental data since U$^{\rm VI}$ was found in niobates, however, the contradiction can be explained by the fact that U$^{\rm VI}$ in practice exists not in a perfect crystal phase, but in the metamict one.

Additionally, since the really synthesized YNbTiO$_6$ structures can not be considered as perfect (periodic) crystals because the Nb and Ti atoms are statistically distributed within them occupying the same Wyckoff positions, different Ti $\leftrightarrow$ Nb substitutions are studied and corresponding structural changes are estimated.

Though all the calculations in the present study are performed within the DFT framework, the CTEP method, being applied to the crystal fragments of moderate sizes, allows one to use more sophisticated wave-function based methods, which we are going to apply to the niobate crystals in our future studies similar to that in xenotime\cite{oleynichenko2023compoundtunable}.
	
%=========================
\section{Acknowledgements}

The present study was supported by the Russian Science Foundation under grant no.~20-13-00225, https://rscf.ru/project/23-13-45028/.

  \bibliography{JournAbbr,QCPNPI,TitovLib,Maltsev}

\end{document}